\documentclass[aps,prl]{revtex4} 
\usepackage[locale=UK,
number-mode=math,
unit-mode=text,
number-unit-product=\ ,
retain-zero-exponent=true]{siunitx}
\DeclareSIUnit{\pixel}{px}
\DeclareSIUnit{\fps}{fps}

\usepackage{palatino,microtype,pifont,ifplatform}
\usepackage{marvosym,amsmath,amsfonts,wasysym}
\usepackage{graphicx,setspace,eurosym}
\usepackage[hidelinks]{hyperref}
\usepackage{xfrac,amssymb,xcolor}

\definecolor{mygg}{RGB}{0, 128, 0}
\definecolor{myrr}{rgb}{0.695,0.133,0.133}

\graphicspath{{./figs/}}
\graphicspath{{../figurematter/latex/figs/}}

\renewcommand{\d}{\mathrm{d}}
\newcommand{\kindex}[2]{\ensuremath{{#1}_{\scalebox{0.65}{#2}}}}
\begin{document}
\title{Universal translational velocity of vortex rings behind conical objects}
\author{Guillaume de Guyon}
\author{Karen Mulleners}
\affiliation{École polytechnique fédérale de Lausanne, Institute of Mechanical Engineering, Unsteady flow diagnostics laboratory, 1015 Lausanne, Switzerland}
\maketitle
\section*{Abstract}
Ring vortices are efficient at transporting fluid across long distances. They can be found in nature in various ways: they propel squids, inject blood in the heart, and entertain dolphins. These vortices are generally produced by ejecting a volume of fluid through a circular orifice. The impulse given to the vortex rings moving away results in a propulsive force on the vortex generator. Propulsive vortex rings have been widely studied and characterised. After four convective times, the vortex moves faster than the shear layer it originates from, and separates from it. When the vortex separates, the circulation of the vortex reaches a maximum value, and the non-dimensional energy attains a minimum. The simultaneity of these three events obfuscates the causality between them. To analyse the temporal evolution of the non-dimensional energy of ring vortices independent of their separation, we analyse the spatiotemporal development of vortices generated in the wake of cones. Cones with different apertures and diameters were accelerated from rest to produce a wide variety of vortex rings. The energy, circulation, and velocity of these vortices were extracted based on time-resolved velocity field measurements. The vortex rings that form behind the cones have a self-induced velocity that causes them to follow the cone and they continue to grow as the cone travels well beyond the limiting vortex formation times scales observed for propulsive vortices. The non-dimensional circulation, based on the vortex diameter, and the non-dimensional energy of the drag vortex rings converge after three convective times to values comparable to their propulsive counterparts. This result proves that vortex pinch-off does not cause the non-dimensional energy to reach a minimum value. The limiting values of the non-dimensional circulation and energy are mostly independent of the cone geometry and translational velocity and fall within an interval of $\SI{10}{\percent}$ around the mean value. The velocity of the vortex shows only $\SI{6}{\percent}$ of variation and is the most unifying quantity that governs the formation of vortex rings behind cones.

\section*{Introduction}
Vortex rings are ubiquitous phenomena widely observed in nature.
Many sea creatures produce vortex rings to propel themselves efficiently. Squids, scallops and salps eject water through a circular orifice, producing a high velocity vortex ring and thus thrust \cite{Johnson1972,Linden2004}.
Some fish release vortex rings in their wake by oscillating their tail and pectoral fins \cite{Lauder2002}.
Vortex rings are also efficient at transporting fluid. The blood injected in the left ventricle of the heart forms a vortex ring, and any imperfection in the formation process is symptomatic of severe heart disease \cite{Dabiri2009}. Extinguishing powder can be transported on distances superior to $\SI{100}{m}$ to extinguish oil well fires, by shooting a vortex ring along the axis of the burning gusher \cite{Akhmetov1980}.\\
Vortex rings introduced above may be classified as propulsive vortices. They are generated by ejecting fluid through a circular orifice, or around a fin, and move away from the body they originate from. The momentum given to the fluid results in a propulsive force acting on the body. A second family of vortex rings emerges from this classification. Vortices passively generated in the wake of a moving axisymmetric body. We refer to them as drag vortices.
They are involved in slowing down the fall of Dandelions, improving the seeding on long distances \cite{Cummins2018}.
Vortex rings also form in the wake of parachutes when they deploy and can lead to the collapse of the parachute if not properly considered \cite{Higuchi1996}.\\ \\
Vortex rings have been studied numerically and experimentally. The classical apparatus to study propulsive vortices is to push a volume of fluid out of a cylinder with a piston. A shear layer forms at the exit of the cylinder, then rolls up to create the vortex ring. Time is measured in a non-dimensional form $T^*$ as the ratio between the length of fluid pushed by the piston and the diameter of the exit.
The vortex reaches a maximum circulation $\Gamma$ at $T^* \approx 4$, also known as the vortex formation time. This timing is consistently reported for circulation based Reynolds numbers $\Gamma/\nu$ superior to 2000, and for various piston acceleration profiles \cite{Gharib1998}. When the circulation reaches a maximum, the vortex no longer accepts vorticity from the shear layer and this process is referred to as vortex separation.\\ \\
A first explanation to the separation is derived from the Kelvin-Benjamin variational principle: a steadily translating vortex ring is the maximum state of kinetic energy $E$ on a iso-vortical sheet with constant impulse $I$ \cite{Benjamin}.
This approach led to the computation of the energy of the vortex with respect to its impulse and circulation, $E^*=E/\sqrt{I \Gamma^3}$.
The vortex separates when the non-dimensional energy $E^*$ delivered by the piston falls below the energy of a steadily translating vortex ring \cite{Gharib1998}. The limiting non-dimensional energy is consistently found at $0.3 \pm \SI{15}{\percent}$ for circulation based Reynolds numbers above 2000 \cite{Mohseni2001}.
The non-dimensional energy quantifies the vorticity distribution inside the vortex. The lower the value of $E^*$, the more uniform the vorticity distribution is. For a Hill's spherical vortex, $E^*$ has a low value of $0.16$.
When the shear layer starts to roll-up, for $T^*<1$, vorticity is concentrated near the vortex core and $E^*$ has values above one.
As the piston moves, more vorticity accumulated in the vortex and spreads towards the cylindrical symmetry axis, decreasing the value of $E^*$.
There is a practical limit to the spreading of the vorticity, represented by this limit of $E^*=0.3$.
A possible interpretation of the vortex separation emerges from the study of the stability of vortex rings.
Dynamical system analysis \cite{Shariff2006} and perturbation response of vortices from the Norbury family \cite{Ofarrell2012} showed that vorticity close to the axis of symmetry gets shed in the tail of the vortex, where the shear layer connects to the vortex. This could prevent the accumulation of additional vorticity by the vortex.
\\ \\
The second explanation to the vortex separation is based on a kinematic argument. The vortex separates when it is travelling faster than the shear layer. The velocity of the shear layer is usually estimated to half of the piston velocity \cite{Gharib1998b}. For $E^*\approx 0.3$ the critical separation velocity was calculated at $\SI{59}{\percent}$ of the piston velocity \cite{Shusser2000}.
The translational velocity of the vortex ring depends on its non-dimensional energy.
Saffman~\cite{Saffman1970} estimated the velocity $U_0$ of a viscous steady vortex ring by the relation
\begin{equation}
	E=2U_0I-\dfrac{3}{8}D_0\Gamma^2
\end{equation}
with $D_0$ being the diameter of the vortex ring. This equation is equivalent to
\begin{equation}
	U_0 = \dfrac{\Gamma}{\pi D_0}\left( E^* \sqrt{\pi} +\dfrac{3}{4} \right).
	\label{eq:uo}
\end{equation}
Both energetic and kinematic explanations to the vortex separation are connected and it is not obvious to assess the causality between vorticity spreading, vortex velocity and vortex separation.\\ \\
The tools developed to study propulsive vortices have not been as extensively applied to the analyse of drag vortices.
An apparent reason is that drag vortices do not separate in the same time scale.
Drag vortex rings are usually studied by accelerating a flat plate or a disk in a fluid \cite{Ringuette2007,Fernando:2016du}.
During the first convective times, the vortex created in the wake of the disk develops in a similar way as propulsive vortices.
Around $T^*=4$ the circulation starts increasing at a lower rate, but does not converge, and is not followed by vortex separation \cite{Johari2002}.
For diameter based Reynolds numbers ranging from \numrange{1600}{4000}, the non-dimensional energy decreases down to values between $0.28$ and $0.35$ \cite{Yang2012}.
Later, passed $T^*=10$, azimuthal instabilities break the axisymmetry and lead to separation of the vortex.\\ \\
We propose to experimentally extend the analysis performed on propulsive vortices to drag vortices.
Cones of different apertures, diameters and velocities will be translated to produce a wide variety of vortex rings in their wake. The non-dimensional energy will be measured to assess if it converges to a lower limit when the vortex stays close to its feeding shear layer.
This information will help to understand the causality between vortex separation, vortex velocity and vorticity distribution in a propulsive vortex ring.
A scaling of the vortex circulation, energy and velocity will be performed to identify the most relevant parameters among the geometry and the kinematics of both the cone and the vortex.
\section*{Experiment}
\begin{figure}
\includegraphics{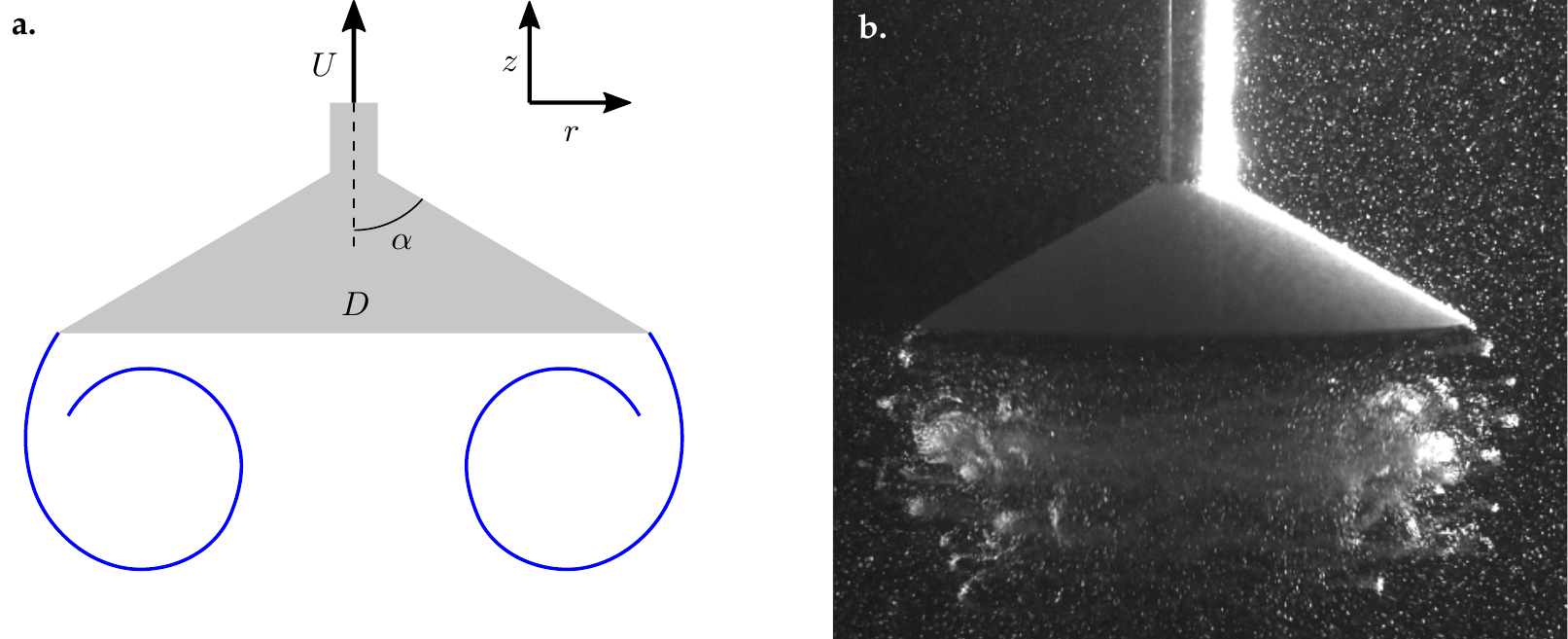}
\caption{\textbf{a} Sketch of the cone and the ring vortex generated in its wake indicating the relevant experimental parameters: cone diameter $D$, cone aperture $\alpha$, constant final cone translation velocity $U$.
The direction of $U$ indicates the upward motion of the cone in the quiescent volume of water.
\textbf{b} Raw image of the seeding particles accumulating in the main vortex ring and highlighting the small scale shear layer vortices.
This raw image was taken after seeding particles had settled on the cone and was not used for processing.
}
\label{fig:setup}
\end{figure}

A cone is immersed in water and translated along its axis of symmetry (\autoref{fig:setup}a).
The translation is performed by a belt driven linear actuator, powered by a NEMA 17 stepper motor.
The cones are 3D printed, their diameters $D$ range from \SIrange{3}{9}{\centi\meter} and their aperture $\alpha$ from $\ang{30}$ to $\ang{90}$.
They are accelerated at $\SI{3}{\meter\per\second\squared}$ from rest, up to velocities $U$ ranging from \SIrange{0.35}{0.70}{\meter\per\second}.
The duration of the acceleration phase varies from \SIrange{0.12}{0.22}{\second} between the fastest and the slowest moving cones.
A series of \num{18} experiments was conducted, with diameter based Reynolds numbers $Re^{}_D=UD/\nu$ comprised between $\num{1e4}$ and $\num{6e4}$.

Particle image velocimetry (PIV) is carried out in a symmetry plane in the wake of the cone during its translation.
The water is seeded with $\SI{30}{\micro\meter}$ aluminium oxide particles which are illuminated with two light emitting diodes (LED).
\autoref{fig:setup}b shows a raw image from one of the cameras, which clearly pictures concentrations of seeding particles indicating small-scale shear layer vortices that accumulate into a large scale coherent ring vortex.
The high concentration of particles in this image is the result of the deposit of the particles on the cone during a break in the measurements.
This image serves as a beautiful visualisation of the main vortex ring and the small scale shear layer vortices that feed into it but is not suitable for PIV analysis.
Only measurement series with homogenous seeding particle distributions have been considered for further analysis.

A field of view of \SI{18x36}{\centi\meter} is recorded at $\SI{1000}{fps}$ by two high speed cameras, each recording one half of the cone trajectory with a definition of \SI{1024x1024}{px}.
The cameras are positioned such that the fields of view touch but do not overlap.
The images are processed with a multi-grid algorithm and a final interrogation window size of \SI{24x24}{px} with an overlap of \SI{60}{\percent}, producing a grid of \num{100x200} velocity vectors with a physical resolution of \SI{1.8}{\milli\meter} between adjacent vectors.
This vector spacing corresponds to \SI{6}{\percent} of the smallest cone diameter.

The length $L$ traveled by the cone is recorded and used here to define the non-dimensional timing of the experiment: $T^*=L/D$.
The cones reach their maximum velocity $U$ after $T^*=0.22$ for large slow cones and after $T^*=2.34$ for small fast cones.
For reference, the non-dimensional time at which the vortex properties converge is around $T^*\approx 3$.
Particle image velocimetry gives access to the velocity field $(u,v)$, from which the vorticity field $\omega$ is derived.
The stream function $\psi$, later used to compute the energy of the vortex, is obtained by integrating the Cauchy-Riemann relations in cylindrical coordinates:
\begin{equation}
	u=\dfrac{1}{r} \dfrac{\partial \psi}{\partial r} \quad , \quad v=-\dfrac{1}{r} \dfrac{\partial \psi}{\partial z}.
	\label{eq:stream}
\end{equation}

\section*{Results}
\subsection*{Spatial and temporal development of a drag vortex ring}
The results presented in this section focus first on the detailed analysis of a single experiment, for a cone of aperture $\alpha=\ang{45}$, diameter $D=\SI{6}{\centi\meter}$ and velocity $U=\SI{0.5}{m.s^{-1}}$. The corresponding Reynolds number $Re^{}_D=\num{3e4}$ is in the middle range of the study. Results from all the experiments are compared in the next section.\\ \\
When the cone is pulled up, fluid moves around it and high velocity gradients raise at the tip of the cone. A shear layer forms and rolls up behind the cone, creating a vortex ring in the wake. With a diameter based Reynolds number superior to \num{1e4}, the shear layer breaks into small scale vortices ($\sim \SI{1}{\milli\meter}$) due to a Kelvin-Helmholtz instability.
The smaller vortices are directly visible on the raw image from the camera (\autoref{fig:setup}b), but are smaller than the output PIV resolution of \SI{1.8}{\milli\meter}.
These instabilities have no effect on the accumulation of vorticity in the vortex \cite{Rosi2017,Mohseni2001}.
\begin{figure}
\includegraphics{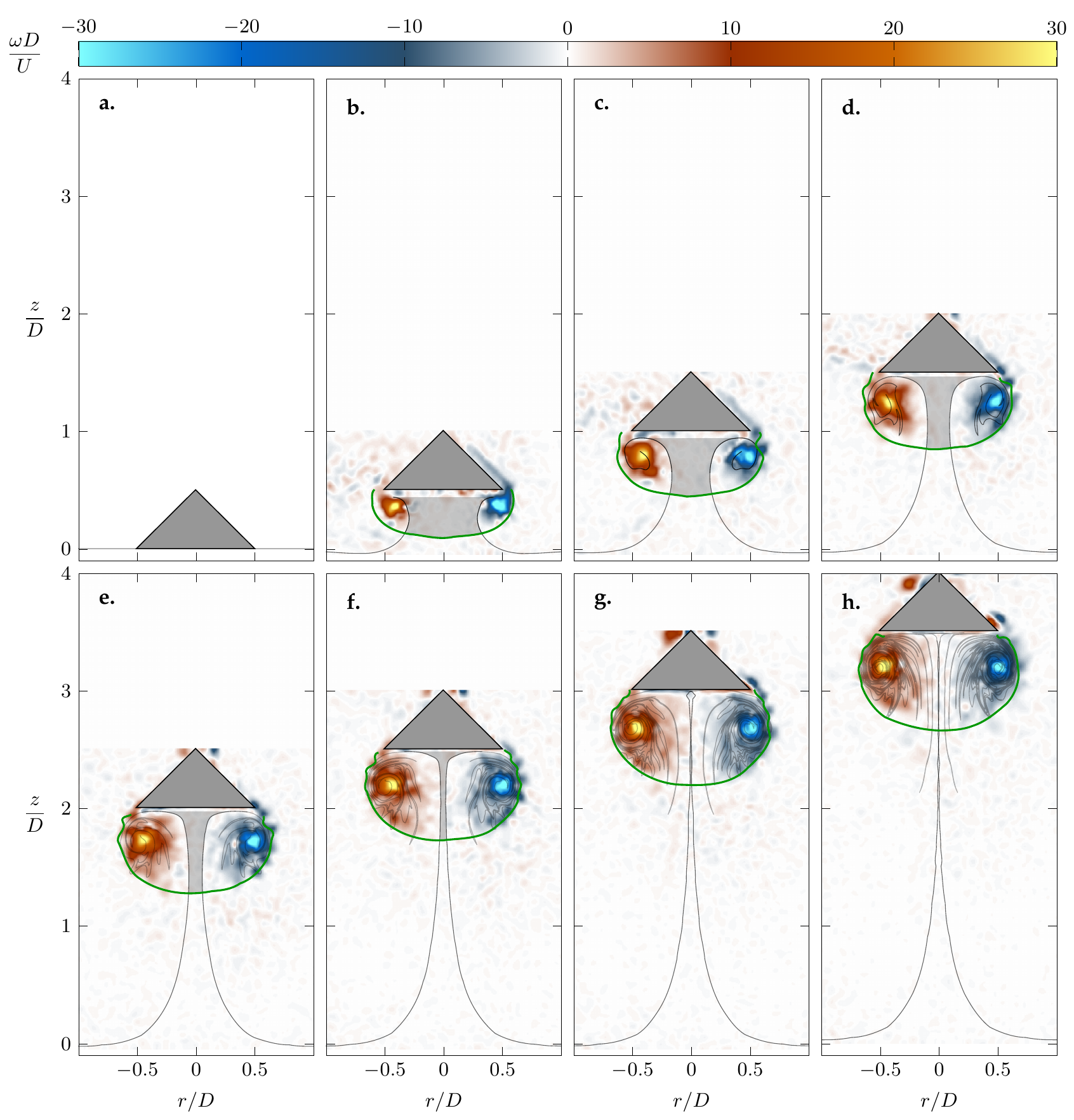}
\caption{Growth of a vortex ring in the wake of a translating cone of $D=\SI{6}{\centi\meter}$, $\alpha=\ang{45}$, $Re^{}_{D}=\num{3e4}$.
The green line delimits the contour of the vortex based on FTLE.
The light grey area from \textbf{b} to \textbf{f} highlights the non-vortical volume of fluid in the vortex and outlines the spreading of vorticity over time.}
\label{fig:vortex_closure}
\end{figure}

\begin{figure}
\includegraphics{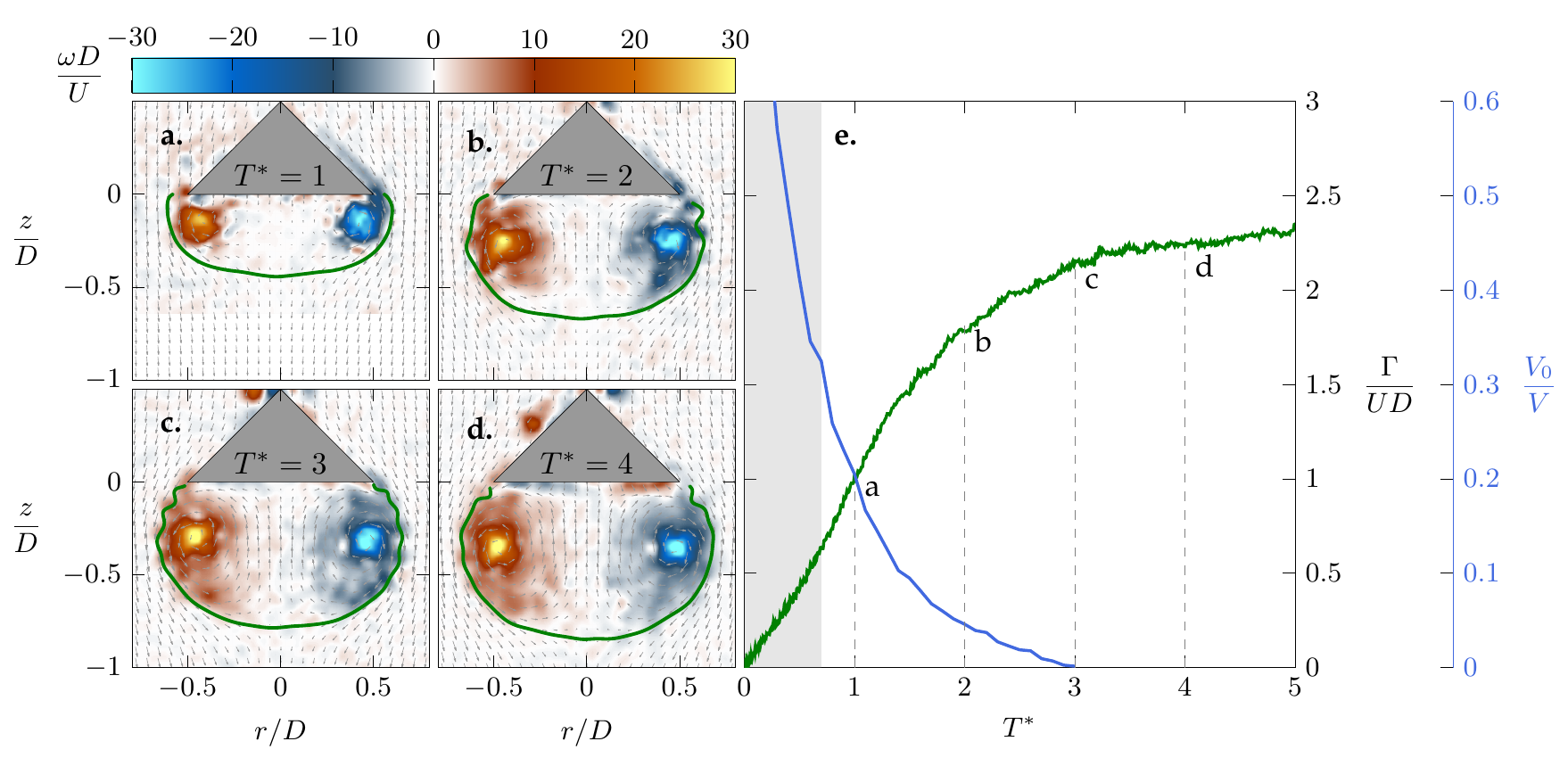}
\caption{Vorticity field and vortex boundary in the wake of a translating cone with $\alpha=\ang{45}$, $D=\SI{6}{\centi\meter}$ and $U=\SI{0.5}{m.s^{-1}}$ at \textbf{a}~$T^*=1$, \textbf{b} $T^*=2$, \textbf{c} $T^*=3$, and \textbf{d} $T^*=4$.
\textbf{e} Temporal evolution of the non-dimensional vortex circulation $\Gamma/UD$ and the volume of non-vortical fluid $V_0$ relative to the total vortex volume $V$.
The grey area indicates the acceleration phase of the cone.}
\label{fig:vor_ftle}
\end{figure}

The vortex contour is determined using Lagrangian methods.
A finite time Lyapunov exponent (FTLE) analysis has been applied to accurately delimit vortices produced by various vortex generators \cite{Shadden2006,OFarrell2010,Green2011,Krishna2018}.
The vortex is delimited by the positive FTLE ridge and the base of the cone (\autoref{fig:vortex_closure}).
This contour does not only contain vortical fluid but also entrained fluid with low or zero vorticity.
Non-vortical fluid is pulled along in the wake of the cone during the initial acceleration from rest.
This non-vortical fluid initially sits below the cone, delimited by the virtual line on \autoref{fig:vortex_closure}a. The integration of this line trajectory delimits the volume of non-vortical fluid from the volume of vortical fluid injected in the vortex at the tip of the cone. The non-vortical volume of fluid, indicated by light grey area on \autoref{fig:vortex_closure}b-f, is entrained and progressively mixed with the vortical fluid.
At $T^*=0.5$ (\autoref{fig:vortex_closure}b), the non-vortical fluid accounts for $\SI{38}{\percent}$ of the volume of the vortex.
At $T^*=3$ (\autoref{fig:vortex_closure}g) there is no more non-vortical fluid.
Vorticity has then fully spread throughout the vortex. \\ \\
The consequences of the spreading of the vorticity is analysed by calculating the vortex circulation $\Gamma=\iint \omega \d r \d z$.
The surface integration is performed on each side of the $r=0$ axis, on the domain delimited by the FTLE contour and the cone base.
The absolute values obtained on both sides of the symmetry axis are averaged to obtain a final circulation value.
The non-dimensional circulation $\Gamma/UD$ increases up to values around $2.3$ (\autoref{fig:vor_ftle}e).
The growth rate of the circulation $\dot{\Gamma}$ decreases progressively.
Around $T^*=1$, a maximum growth rate $\dot{\Gamma}=1.2U^2$ is reached. After $T^*=3$ the rate stabilises around $\dot{\Gamma}=0.06U^2$.
The stabilisation starts when the non-vortical volume of fluid $V_0$ inside the vortex volume $V$ vanishes.

\begin{figure}
\includegraphics{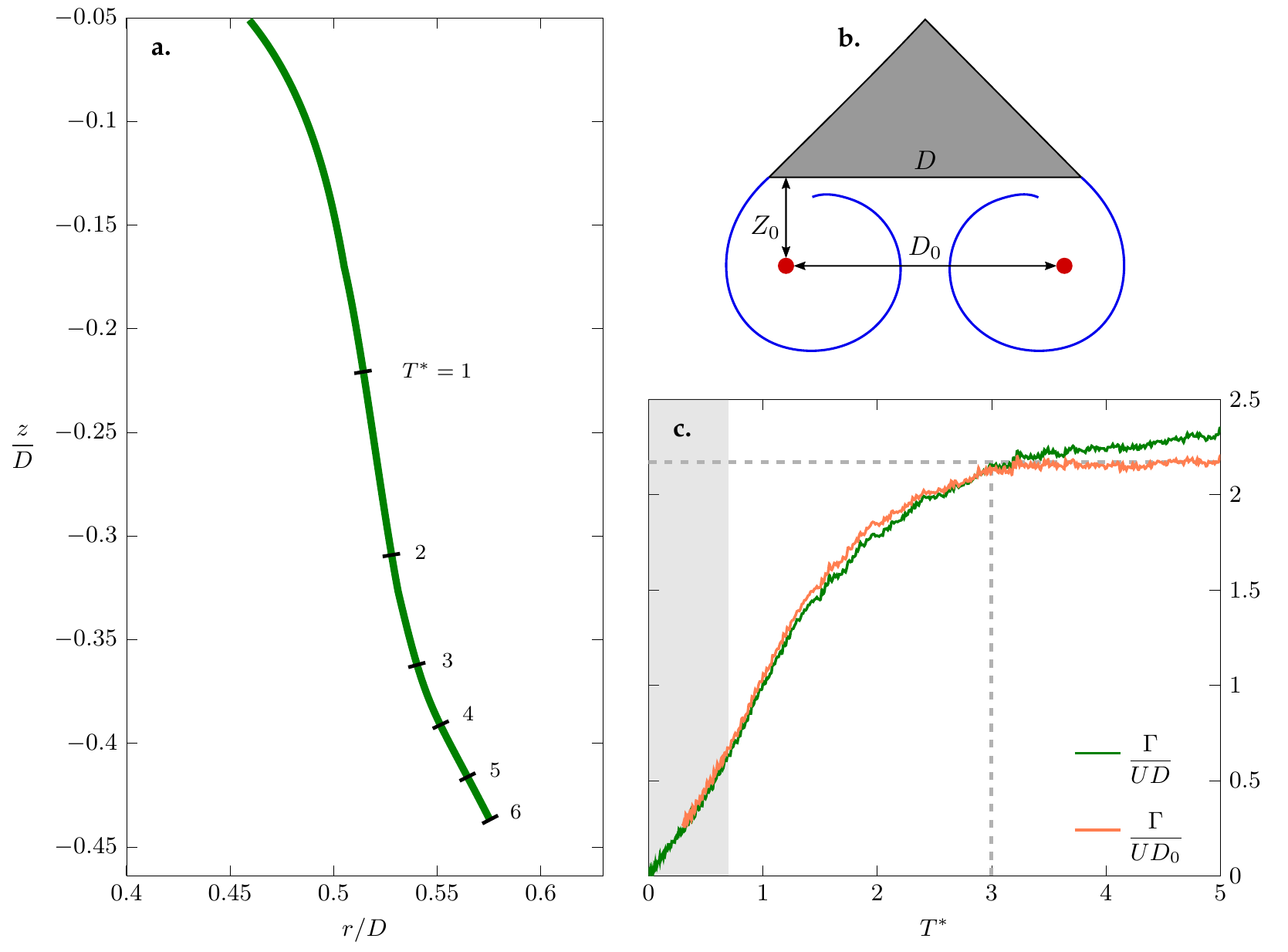}
\caption{\textbf{a} Trajectory of the vortex centre relative to the cone for $\alpha=\ang{45}$, $D=\SI{6}{\centi\meter}$ and $U=\SI{0.5}{m.s^{-1}}$.
\textbf{b} Parametrisation of the vortex centre.
\textbf{c} Circulation of the vortex, non-dimensionalised using the cone diameter $D$ and the vortex diameter $D_0$.
The grey area indicates the acceleration phase of the cone.}
\label{fig:vortex_centers}
\end{figure}

Similar observations where made on vortex rings generated by piston cylinders: at $T^*=4$, the circulation reaches a maximum $\Gamma/UD \approx 2.3$ and the vorticity spreads up to the cylindrical symmetry axis.
The vortex separation for propulsive vortices is attributed to the tail shedding resulting from the vorticity spread \cite{Gharib1998}.
In the present experiment, no separation occurs because the vortex translational velocity is directed towards the cone. The vortex stays in the vicinity of the shear layer, keeps growing and accumulating vorticity after $T^*=3$ (\autoref{fig:vor_ftle}c-d).
\\ \\
The extension of the vortex is quantified by observing the vortex centre $(Z_0,R_0)$, calculated as
\begin{equation}
	Z_0=\frac{\iint \omega zr^2 \d r \d z}{\iint \omega r^2 \d r \d z} ,\qquad R_0^2=\frac{\iint \omega r^2 \d r \d z}{\iint \omega \d r \d z} ,\qquad D_0=2R_0.
	\label{eq:center}
\end{equation}
The integrations are performed on each half of the $r=0$ axis, on the domain delimited by the FTLE contour and the cone base.
The absolute values obtained on both sides of the symmetry axis are averaged to obtain the final values.
Positions are given relative to the base of the cone (\autoref{fig:vortex_centers}b).

\begin{figure}
\includegraphics{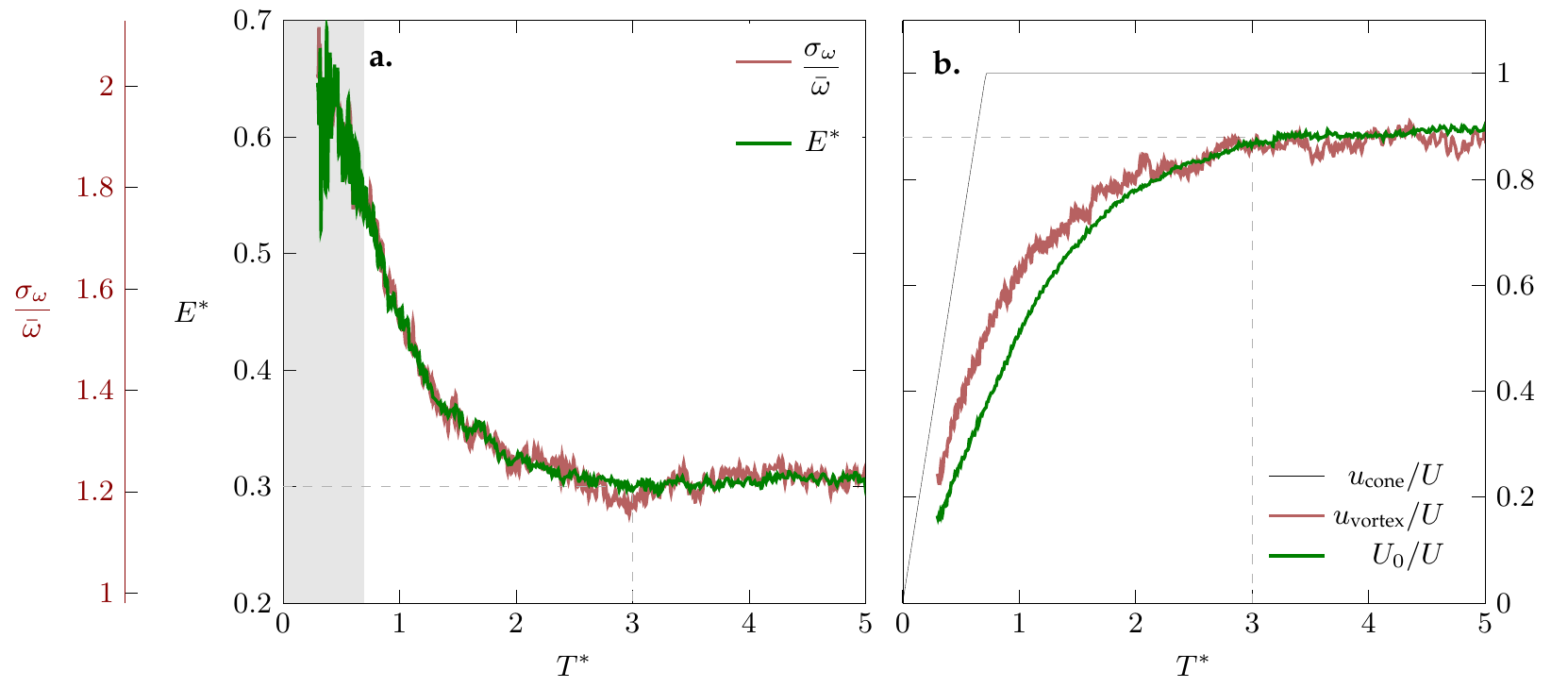}
\caption{\textbf{a} Temporal evolution of the non-dimensional energy and relative vorticity distribution.
The grey area indicates the acceleration phase of the cone.
\textbf{b} Theoretical and measured velocity of the vortex, relative to the maximum cone velocity.
The evolution of the translation velocity of the cone is added for reference.}
\label{fig:Eadim}
\end{figure}

The vortex centre (\autoref{fig:vortex_centers}a) quickly moves away from the cone and has travelled a distance of $0.22D$ at $T^*=1$.
It progressively slows down and for $T^*>3$ the vortex centre moves away from the cone and the axis of symmetry at a more constant velocity.
A distance of $0.03D$ is covered between $T^*=4$ and $T^*=5$.
The increase of the vortex diameter suggests that the cone diameter is not the most suitable parameter for non-dimensionalisation of the circulation.
Using the vortex diameter $D_0$, the non-dimensional circulation reaches a maximum value of $2.2$ at $T^*=3$ and stays constant, whereas the non-dimensional circulation based on $D$ continues to grow (\autoref{fig:vortex_centers}b).
The term $\Gamma/D_0$ has the dimension of a velocity and is featured in \autoref{eq:uo}.
It quantifies the influence of the vortex circulation and dimension on the velocity of the ring.
\\ \\
The other parameter that influences the velocity of the vortex ring is the non-dimensional energy of the vortex $E^*$.
It represents the energy $E$ relative to impulse $I$ and circulation $\Gamma$:
\begin{equation}
	\quad I=\pi \iint \omega r^2 \d r \d z \quad,\quad E= \pi \iint \psi \omega \d r \d z \quad,\quad E^*=\dfrac{E}{\sqrt{I \Gamma^3}} \quad .
	\label{eq:int}
\end{equation}
The non-dimensional energy quantifies the distribution of the vorticity inside the vortex.
It is compared to a more statistical definition of the vorticity distribution: the standard deviation of the vorticity $\sigma_{\omega}$ relative to its average $\bar{\omega}$. The evolution of $E^*$ and $\sigma_{\omega}/\bar{\omega}$ are presented in \autoref{fig:Eadim}a.
The non-dimensional energy and relative vorticity distribution have a similar evolution and an empirical relation $\sigma_{\omega}/\bar{\omega}=2.3E^*+0.5$ can be derived.
Although this relation is specific to this experiment, it confirms that $E^*$ is a valid quantifier of vorticity distribution.
During the first convective time, the shear layer starts to roll-up and vorticity is concentrated in the vortex core.
Non-dimensional energy has high values around $0.7$. As vorticity keeps accumulating in the vortex, $E^*$ continuously drops towards a limiting value of $0.3$, reached at $T^*=3$.
A minimal value between $0.27$ and $0.35$ was also observed for vortex rings generated by piston cylinders \cite{Gharib1998,Danaila2018,Nitsche2001} and corresponds to the moment when the vortex separates from its feeding shear layer.
This limit on $E^*$ answers one of the questions that motivated this experiment: the non-dimensional energy of a vortex ring does not decrease further, even if it stays connected to its feeding shear layer. Contrary to vortex rings generated by piston cylinders, vorticity still accumulates in the vortex at a low rate, without noticeable effect on the vorticity distribution.\\ \\
The quantities $\Gamma/D_0$ and $E^*$ characterise respectively the overall vorticity present in the vortex and the distribution of the vorticity in the vortex.
Both reach a limiting value at $T^*=3$.
As a consequence, the theoretical translational velocity $U_0$ of the vortex ring, derived from Saffman \cite{Saffman1970}, also reaches a steady value:
\begin{equation}
	U_0 = \dfrac{\Gamma}{\pi D_0}\left( E^* \sqrt{\pi} +\dfrac{3}{4} \right).
	\label{eq:u1}
\end{equation}
The theoretical velocity (\autoref{eq:u1}) is compared with the measured vortex velocity $\kindex{u}{vortex}$, obtained by averaging the axial velocity component inside the vortex volume.
Both theoretical and measured velocities converge to a value of $0.9U$ (\autoref{fig:Eadim}).
The theoretical velocity underestimates the measured velocity during the transient phase, due to the fact that \autoref{eq:u1} is only valid for a steady vortex ring.

Since the vortex stays attached to the cone, the velocity deficit of the vortex compared to the cone should not be interpreted as a vortex separation, but rather as an increase of the vortex volume.
\subsection*{Scaling of the vortex characteristics}
\begin{figure}
\includegraphics{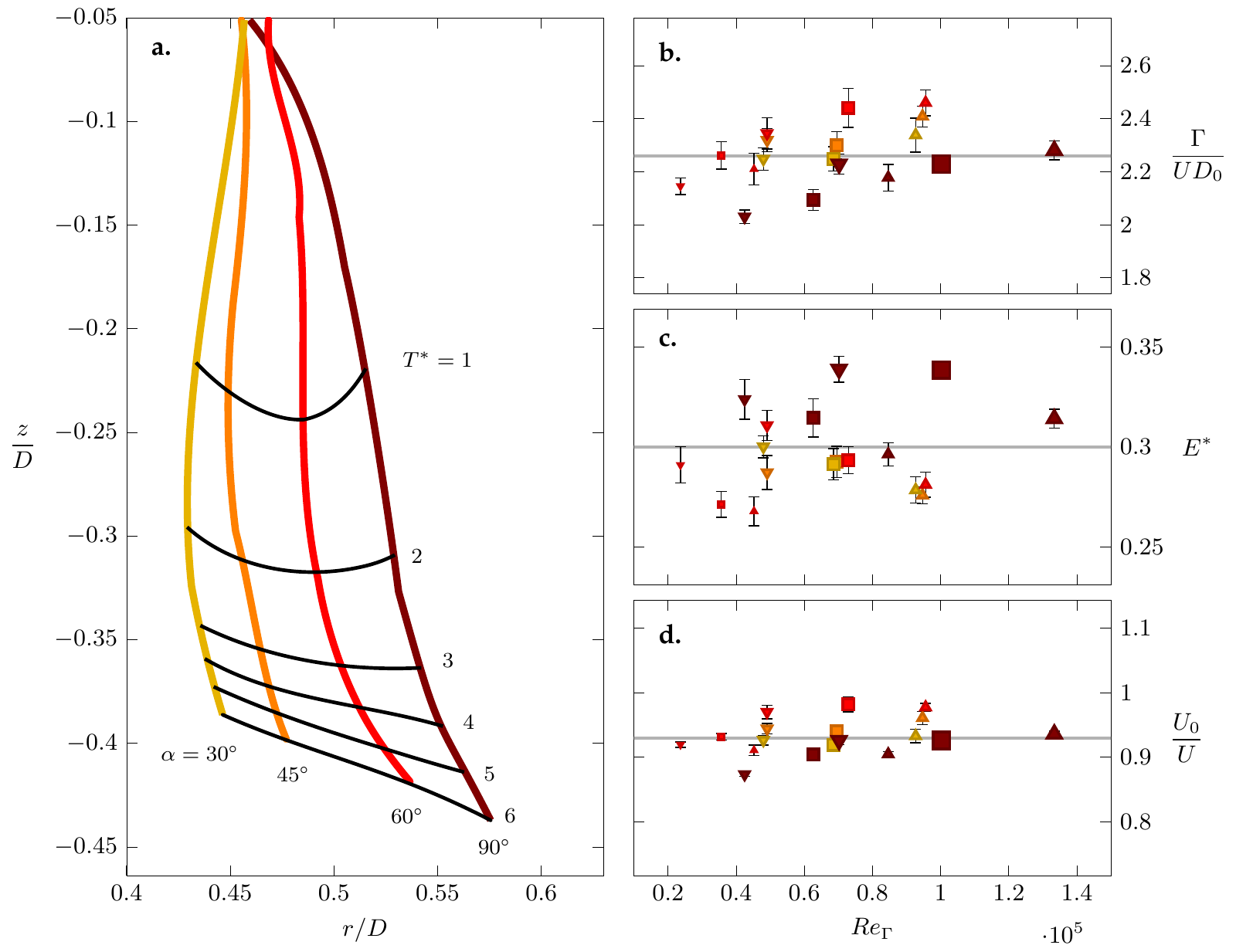}
\caption{\textbf{a} Vortex centre trajectories for 4 aperture angles, at $D=\SI{6}{\centi\meter}$ and $U=\SI{0.5}{\meter\per\second}$.
\textbf{b-d} Maximum circulation, minimum non-dimensional energy, and maximum theoretical vortex velocity for all experiments.
The grey lines indicate the average value across all parameter variation.
The error bars reflect the uncertainty on the converged values quantified as explained in the \hyperref[appendix:b]{appendix}.
The shape of the markers indicate the final translation velocity of the cone
$\triangledown$: $U=\SI{0.35}{m.s^{-1}}$, $\square$: $U=\SI{0.5}{m.s^{-1}}$, $\vartriangle$: $U=\SI{0.65}{m.s^{-1}}$.
The size of the markers reflects the diameter of the cone $\circ$: $D=\SI{3}{\centi\meter}$, $\Circle$: $D=\SI{6}{\centi\meter}$, $\bigcirc$: $\SI{9}{\centi\meter}$.
}
\label{fig:comp}
\end{figure}

The influence of the cone geometry and kinematics on the vortex circulation, non-dimensional energy and velocity are analysed in this section.
A discussion of the influence of the vortex detection method on the integral values and an estimation of the experimental uncertainty is presented in the appendix.
All experiments are sorted by their circulation based Reynolds number.
As seen in the previous section, the circulation does not reach a clear maximum, but $\Gamma/D_0$ does. We therefore introduce a Reynolds number $Re^{}_{\Gamma}=\dfrac{\Gamma D}{\nu D_0}$ based on the maximum value of $\Gamma/D_0$.\\ \\
Different cone aperture were used, from $\alpha=\ang{30}$ to $\alpha=\ang{90}$ (flat disk).
The trajectory of the vortex centre is extracted for four cones of different aperture and identical diameter $D=\SI{6}{\centi\meter}$ and velocity $U=\SI{0.5}{\meter\per\second}$ (\autoref{fig:comp}). The four trajectories show an increase of the vortex diameter passed $T^*=2$.
A larger cone aperture leads to a larger vortex diameter because the cone deviates the flow more in the radial direction.
At $T^*=3$ the vortex radius ranges from $D_0=0.88D$ at $\alpha=\ang{30}$ to $D_0=1.1D$ at $\alpha=\ang{90}$.

The maximum non-dimensional circulation $\Gamma/UD_0$ is calculated for all cases and presented in \autoref{fig:comp}b.
All values presented in \autoref{fig:comp}b-d are obtained after the cone has reached its final constant velocity.
The average value is \num{2.26}, with variations of $\pm \SI{9}{\percent}$.
Cones with an aperture of $\ang{90}$, or disks, exhibit lower non-dimensional circulations.
For structural reason they are not strictly disks: they have a thickness of $0.06D$ and a reverse sweep of $\ang{30}$. It is believed to be responsible for the lower non-dimensional circulation of the disk compared to the cones.
The effect is reversed for the non-dimensional energy. Disks have slightly higher values (\autoref{fig:comp}c). Non-dimensional energy converges in average to a minimum of $0.3$, with variations of $\pm \SI{12}{\percent}$.\\ \\
The non-dimensional circulation and energy are largely independent on the Reynolds number.
Similar behaviour is observed for vortices generated by piston cylinders. For various simulations with $Re^{}_{\Gamma}>2000$, non-dimensional circulation and energy were recorded to have variations of respectively $\pm \SI{10}{\percent}$ and $\pm \SI{15}{\percent}$ \cite{Mohseni2001}.\\ \\
From the circulation and the non dimensional energy, the theoretical velocity $U_0$ of the vortex ring is computed (\autoref{eq:u1}).
The relative velocity $U_0/U$ has an average of $0.93$, within an interval of $\pm \SI{6}{\percent}$ (\autoref{fig:comp}d).
The interval of variation is smaller than the ones of $\Gamma/UD_0$ ($\pm \SI{9}{\percent}$) and $E^*$ ($\pm \SI{12}{\percent}$).
The variations in non-dimensional circulation and non-dimensional energy compensate to produce a regular relative velocity.
In particular the disks, which have higher non-dimensional energy and lower circulation, have a relative velocity equivalent to the cones.
This suggests that the cone's velocity is a better scaling parameter than the non-dimensional circulation or energy.

\section*{Conclusion}
A vortex ring produced by a piston cylinder simultaneously separates, reaches a minimal non-dimensional energy, and outpaces its feeding shear layer.
The simultaneity of these three events obfuscate the causality between them.
To analyse the temporal evolution of the non-dimensional energy of ring vortices independent of their pinch-off, we focused on vortices generated in the wake of cones.
Cones with different apertures and diameters were accelerated from rest to produce a wide variety of vortex rings.
The initial development and growth of these vortex rings were studied experimentally using time-resolved velocity field measurements.
\\ \\
The vortex rings that form behind the cones have a self-induced velocity that cause them to follow the cone beyond the typical vortex formation time scales observed for vortex rings emanating from a piston cylinder apparatus.
Another difference is that propulsive vortex rings reach a maximum in circulation but the circulation of drag vortex rings keeps increasing and does not reach a clear plateau.
However, for $T^*>3$, the circulation of drag vortices increases proportionally to the vortex size and the circulation non-dimensionalised by the vortex diameter $\Gamma/UD_0$ converges to a limiting value between \num{2.05} and \num{2.45}.
The non-dimensional energy is linearly related to the relative standard deviation of the vorticity, demonstrating that $E^*$ is a measure of vorticity distribution inside the vortex ring.
At the start of the vortex formation, vorticity is concentrated near the vortex core and $E^*$ is at its highest.
In time, the vortex grows and the vorticity distribution spreads, which is reflected by a decrease in $E^*$.
The non-dimensional energy converges around $T^*=3$ to a minimum value between \num{0.27} and \num{0.35}.
Similar values of non-dimensional circulation and energy were observed for vortices produced by piston cylinders.
This results proves that vortex pinch-off does not cause the non-dimensional energy to converge to a minimum value.\\ \\
The non-dimensionalised energy, circulation, and velocity of the ring vortices reach constant values independent of the cone diameter, aperture angle, and translational velocity when scaled based on the vortex diameter instead of the cone diameter.
The limiting values of the circulation and the energy experience variations of $\SI{9}{\percent}$ and $\SI{12}{\percent}$ across the tested parameter range.
These variations compensate each other to produce a constant vortex velocity of $U_0 = 0.93U \pm \SI{6}{\percent}$.
The difference between the vortex velocity and the cone velocity does not indicate vortex separation but is the result of the spatial growth of the vortex.
The vortex velocity is the most unifying quantity to scale and predict the development of vortex rings behind various cone geometries.

\appendix
\section{Appendix}
\subsection{Sensitivity to the vortex detection method}
\begin{figure}
	\includegraphics{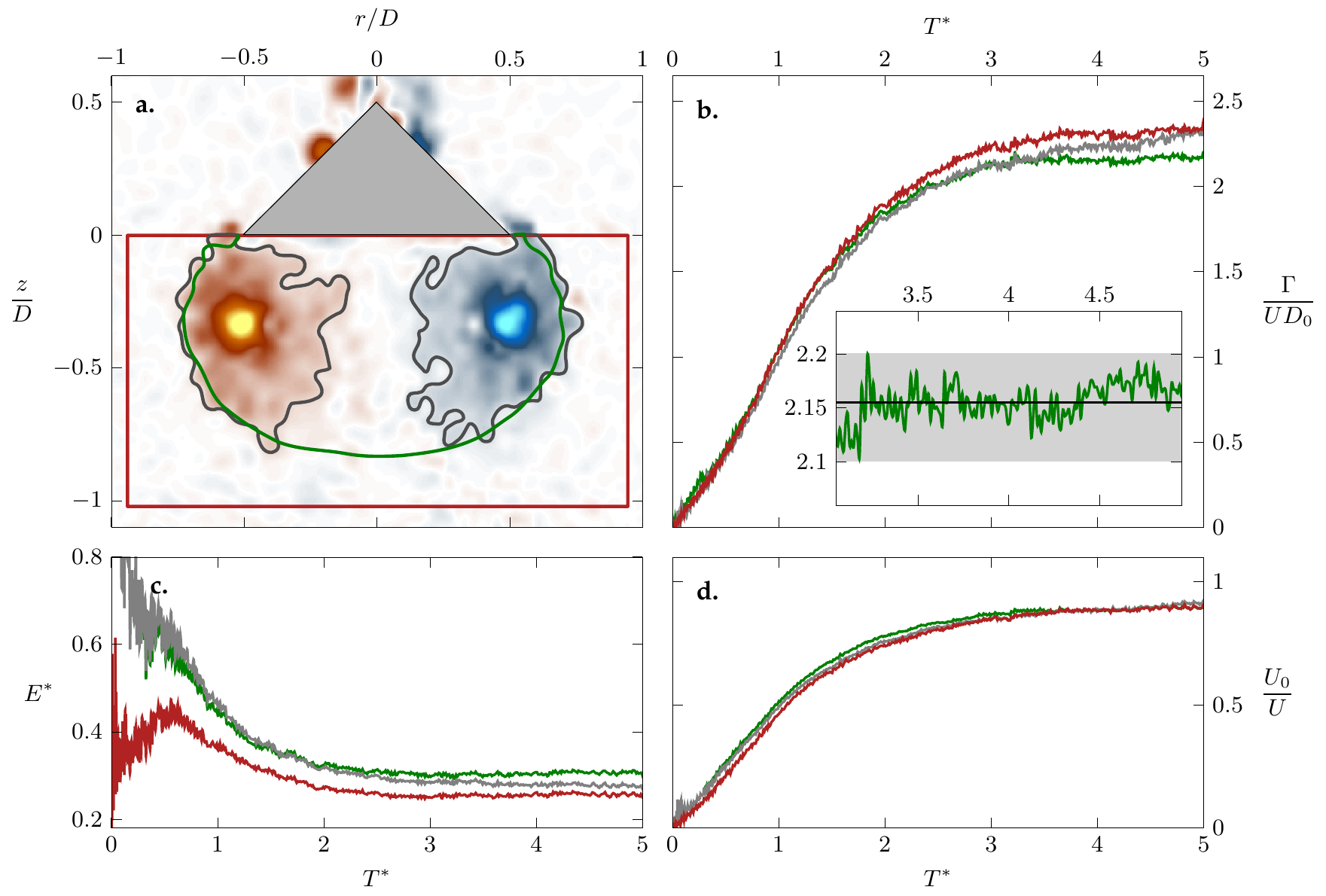}
	\caption{\textbf{a} Vorticity field at $T^*=3$ and contours used to compute vortex characteristics including the FTLE ridge (\textbf{\textcolor{mygg}{---}}),
	isocontour of vorticity $|\omega D/U |=1$ (\textbf{\textcolor{gray}{---}}),
	and a rectangular box (\textbf{\textcolor{myrr}{---}}).
	\textbf{b} Temporal evolution of the non-dimensional circulation, \textbf{c} non-dimensional energy, and \textbf{d} theoretical vortex velocity relative to the cone velocity.
The inset in \textbf{b} zooms in on the convergence for the circulation calculated within the FTLE boundary.
The black line indicates the average value between $T^*=3$ and $T^*=5$ and is considered as the converged circulation result.
The amplitude of the variations are outlined by the grey rectangle and this is considered as the uncertainty for the measured circulation values.
}
\label{fig:appendix}
\end{figure}

To estimate the uncertainty of the scaling values computed in this article (\autoref{eq:center}-\ref{eq:int}) and their sensitivity to the selection of the integration contour, we compare here the results for three different integration contours.
The contour that is used for the final results presented in \autoref{fig:comp} is the FTLE contour.
In addition to the FTLE contour, we consider here the iso-vorticity contour $|\omega D/U |=1$, and a rectangular contour of dimensions $1.8D \times D$ (\autoref{fig:appendix}\textbf{a}).
The iso-vorticity contour is slightly wider than the FTLE one and contains also vorticity patches outside of the FTLE contour that will be convected into the wake.
The rectangular box is the easiest contour to use, but it will also contain vorticity that should not be considered as part of the vortex.
The rectangular and vorticity contours both contain more vorticity that the FTLE contour encloses, resulting in higher circulations and larger vortex diameters.
At $T^*=4$, the non-dimensional circulation is respectively $\SI{7}{\percent}$ and $\SI{4}{\percent}$ higher than with the FTLE method (\autoref{fig:appendix}\textbf{b}).
The effect is reversed for the non-dimensional energy, as the vorticity added by the larger contours is of low value and more uniform than in the vortex core (\autoref{fig:appendix}\textbf{d}).
For the rectangular and the iso-vorticity contour, $E^*$ is respectively $\SI{16}{\percent}$ and $\SI{9}{\percent}$ lower.
Variations in circulation and non-dimensional energy compensate in the computation of the velocity (\autoref{fig:appendix}\textbf{c}).
Both the rectangular and the iso-vorticity contour result in the same converged velocity value within $\SI{3}{\percent}$ of the one computed on the FTLE contour.
The standard deviation of the bias introduced by the choice of the integration contour is $\SI{1.5}{\percent}$ for the tested parameter space and can be considered constant and does not affect the conclusion of the paper.

\section{Accuracy of the convergence}
\label{appendix:b}
The non-dimensional circulation, energy, and vortex velocity presented in \autoref{fig:comp}b-d are all extracted from their converged values.
The converged value is obtained by averaging the quantity between $T^*=3$ and $T^*=5$ (\autoref{fig:appendix}b).
The full range of variations measured between $T^*=3$ and $T^*=5$ determines the uncertainty on the convergence value and is represented by the error bars in \autoref{fig:comp}b-d.
The maximum uncertainty is $\SI{3}{\percent}$ for the non-dimensional circulation, $\SI{4}{\percent}$ for the non-dimensional energy, and $\SI{1}{\percent}$ for the relative velocity of the vortex.
\bibliography{ms2}
\bibliographystyle{abbrv}
\end{document}